# A Fast Sub-pixel Motion Estimation Algorithm for H.264/AVC Video Coding

Weiyao Lin[1], Krit Panusopone[2], David M. Baylon[2], Ming-Ting Sun[3], Zhenzhong Chen[4] and Hongxiang Li[5]

[1] Institute of Image Communication and Information Processing, Dept. of Electronic Engineering, Shanghai Jiao Tong University, Shanghai 200240, China

[2] Advanced Technology Department, Mobile Devices & Home, Motorola, Inc., San Diego, CA 92121, USA

[3] Dept. of Electrical Engineering, University of Washington, Seattle, WA 98195, USA

[4] School of Electrical and Electronic Engineering, Nanyang Technological University, Singapore

[5] Dept. of Electrical and Computer Engineering, North Dakota State University, Fargo, ND USA

## Abstract

Motion Estimation (ME) is one of the most time-consuming parts in video coding. The use of multiple partition sizes in H.264/AVC makes it even more complicated when compared to ME in conventional video coding standards. It is important to develop fast and effective sub-pixel ME algorithms since (a) The computation overhead by sub-pixel ME has become relatively significant while the complexity of integer-pixel search has been greatly reduced by fast algorithms, and (b) Reducing sub-pixel search points can greatly save the computation for sub-pixel interpolation. In this paper, a novel fast sub-pixel ME algorithm is proposed which performs a 'rough' sub-pixel search before the partition selection, and performs a 'precise' sub-pixel search for the best partition. By reducing the searching load for the large number of non-best partitions, the computation complexity for sub-pixel search can be greatly decreased. Experimental results show that our method can reduce the sub-pixel search points by more than 50% compared to existing fast sub-pixel ME methods with negligible quality degradation.

## I. Introduction

H.264/AVC is the state-of-the-art video coding standard established by ITU-T and ISO/IEC.

H.264/AVC uses many new techniques and is able to save more than 50% in bitrate while having similar video quality compared to the MPEG-2 video coding standard [1].

Motion Estimation (ME) is one of the most time-consuming parts in video coding. Developing fast algorithms for ME to reduce computational complexity in video coding has been an important and challenging problem. In the H.264/AVC Joint Model (JM) [5], the ME process contains two stages: integer pixel search over a large area and sub-pixel search around the best selected integer pixel. Since H.264/AVC uses 7 partition sizes for inter-frame prediction (16×16, 16×8, 8×16, 8×8, 8×4, 4×8 and 4×4), the complexity of multi-partition ME is high [2]. It is becoming more critical to develop fast and effective sub-pixel ME algorithms for H.264/AVC. Firstly, the computation overhead by sub-pixel ME has become relatively significant while the complexity of integer-pixel search has been greatly reduced by fast algorithms. For example, there have been integer-pixel ME algorithms [4, 10, 16] that only need between 3 and 5 integer search points to calculate the final integer Motion Vector (MV). The computation in the 16-point sub-pixel search method used in the JM thus becomes comparatively large. Secondly, typical sub-pixel searches require interpolating sub-pixel values for computing the Sum of Absolute Difference (SAD). Reducing sub-pixel search points can also reduce the interpolation computation time.

In this paper, a novel sub-pixel ME algorithm is proposed for H.264/AVC, which performs a 'rough' sub-pixel search before the partition selection, and performs a 'precise' sub-pixel search for the best partition. By reducing the searching load for the large number of non-best partitions, the computation complexity for sub-pixel search can be greatly decreased. Experimental results show that the proposed algorithm can significantly reduce the number of sub-pixel search points compared to other fast sub-pixel ME algorithms [6-9], with negligible quality degradation.

The rest of this paper is organized as follows. Section II reviews existing research on sub-pixel ME. Section III provides in-depth analysis on how to further reduce the search points for sub-pixel ME for multiple partitions. The proposed algorithm is described in Section IV. Section V shows the experimental results and Section VI concludes the paper.

## II. Related Work

Chen et al. [6] analyzed the difference between the integer-pixel matching error surface and the sub-pixel matching error surface. According to Chen's analysis, the integer-pixel matching error surface is far from a unimodal surface inside the searching window due to the complexity of the video content. The assumption of unimodal will easily result in trapping in a local minimum. However, for the sub-pixel matching error surface, the unimodal surface assumption holds in most cases because of the smaller search range of sub-pixel ME as well as the high correlation between sub-pixels due to the sub-pixel interpolation.

There has been much research on fast sub-pixel ME [6-9, 17]. Most of these methods are based on the unimodal surface assumption and perform the sub-pixel search in two steps:

1) Predict a sub-pixel MV (*SPMV*), and
2) Perform a small area search around the *SPMV* to obtain the final sub-pixel MV.

The method to get the sub-pixel predicted MV can be summarized in two ways: using spatiotemporal information and modeling the Sum of Absolute Difference (SAD) surface.

Chen et al. [6] and Yang et al. [8] used spatiotemporal information to get the *SPMVs*. In [6], a Center Biased Fractional Pixel Search (CBFPS) fast sub-pixel ME method is studied, where the MVs of neighboring MBs were used to get the *SPMV* as in Eqn (1),

$$SPMV = (pred\_mv - MV)\%\beta \tag{1}$$

where *pred_mv* is the MV prediction of the current partition (in sub-pixel resolution), *MV* is the best integer-pixel MV of the current partition (*β=4* in the 1/4-pixel case and *β=8* in the 1/8-pixel case) and % represents the modulo operation. In [8], a larger partition MV (e.g., 16x8 inter-mode MV takes a 16x16 MV as a reference) or previous frame MV was used to get the *SPMV*. If combined with the *SPMV* from CBFPS, the accuracy of the *SPMV* can be greatly increased.

A more popular way to get the *SPMV* is to use a function (in most cases a second-order function) to model the SAD surface [7, 9]. If the matching errors of the best integer-pixel MV and its neighboring positions are known, the coefficients of the function can be solved. The position that corresponds to the smallest value in the SAD surface is then chosen as the *SPMV*.

Many functions can be used to model the SAD surface. Example second-order functions are listed as Eqns (2) and (3).

$$f(x,y) = c_1x^2 + c_2xy + c_3y^2 + c_4x + c_5y + c_6 \qquad (2)$$

$$f(x,y) = c_1x^2 + c_2x + c_3y^2 + c_4y + c_5 \qquad (3)$$

where $x$ and $y$ are coordinates of the surface, and $f(x, y)$ is the matching error (SAD) value. Normally, the best integer-pixel position is set to be located at $(0, 0)$, so its neighboring integer-pixel positions are at $(1, 0)$, $(-1, 0)$, $(0, 1)$, $(0, -1)$, etc. As the number of model function coefficients increases, more integer-pixel neighboring SADs are needed.

In [7, 9], Eqn (3) was used to determine one of the *SPMV*s, which used the best integer-pixel SAD and the SADs of its four diamond integer neighbors. Given these SAD values, the coefficients of Eqn (3) can be computed. The *SPMV* can then be calculated as:

$$SPMV = (x_p, y_p) = \arg\min_{x,y} f(x, y) = (\frac{-B}{2A}, \frac{-D}{2C}) \qquad (4)$$

where

$$\begin{cases} A = (I+J)/2 \\ B = (I-J)/2 \\ C = (K+L)/2 \\ D = (K-L)/2 \end{cases} \begin{cases} I = f(1,0) - f(0,0) \\ J = f(-1,0) - f(0,0) \\ K = f(0,1) - f(0,0) \\ L = f(0,-1) - f(0,0) \end{cases}, \text{ and } \begin{cases} f(0,0) = SAD(0,0) = c_5 \\ f(1,0) = c_1 + c_2 + c_5 \\ f(-1,0) = c_1 - c_2 + c_5 \\ f(0,1) = c_3 + c_4 + c_5 \\ f(0,-1) = c_3 - c_4 + c_5 \end{cases}$$

If $(x_p, y_p)$ is a fractional vector, its components are quantized into quarter-pixel units.

Furthermore, Xu et al. [17] proposed to use early terminations to further reduce the search points from the CBFPS method.

## III. Analysis on Reducing Sub-pixel Search Points with Multiple Partitions

As shown in Section II, most previous fast sub-pixel ME methods reduce the number of search points by only searching the reduced area around the *SPMV*. For H.264/AVC multiple partition sizes, they attempt to find the 'best' sub-pixel MV (with the smallest SAD) for each partition before the partition selection, as shown in Fig. 1(a).

However, in practice, only the best partition of the MB needs precise sub-pixel MVs. The MVs of other partitions are only used for the inter-mode selection. They are no longer useful after the best partition is selected. If a sub-pixel SAD is good enough to select the best partition, there is no need to search for more precise sub-pixel points in the first stage.

Therefore, if only a 'rough' sub-pixel motion search is performed for each partition (the resulting MV does not necessarily have the smallest SAD), and a 'precise' sub-pixel MV is determined only for the best partition selected, then the number of search points for the non-best partitions can be reduced greatly. As shown in Fig. 1(b), the purpose of the first stage ME is to obtain a rough sub-pixel SAD which is close to the best SAD. The integer-pixel SAD surface information can be used to decide whether the sub-pixel SAD is close to the best one or not. Based on the above discussion, we propose a new Rough-strategy-based Fast Sub-pixel Motion Estimation algorithm (RFSME) described in detail in the next section.

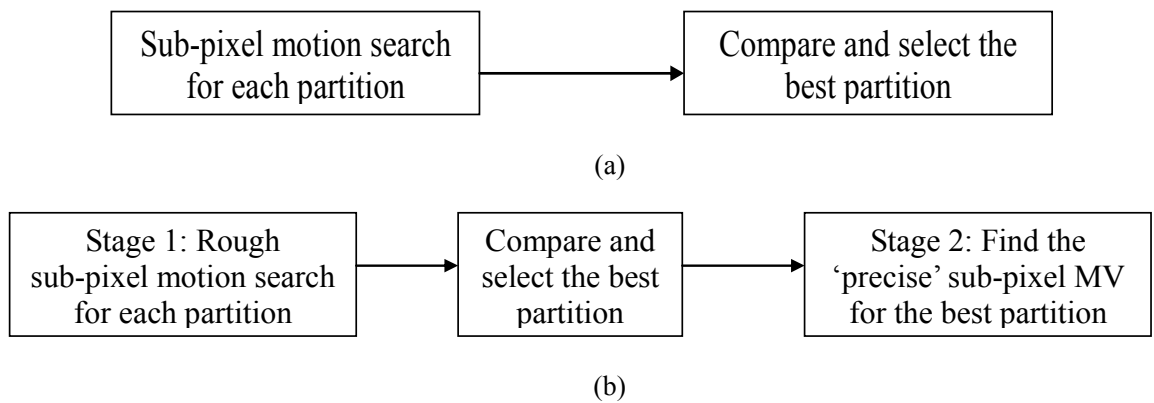

Fig. 1 Fast sub-pixel ME approaches.
(a) Process for previous fast sub-pixel ME methods. (b) Proposed fast sub-pixel ME process.

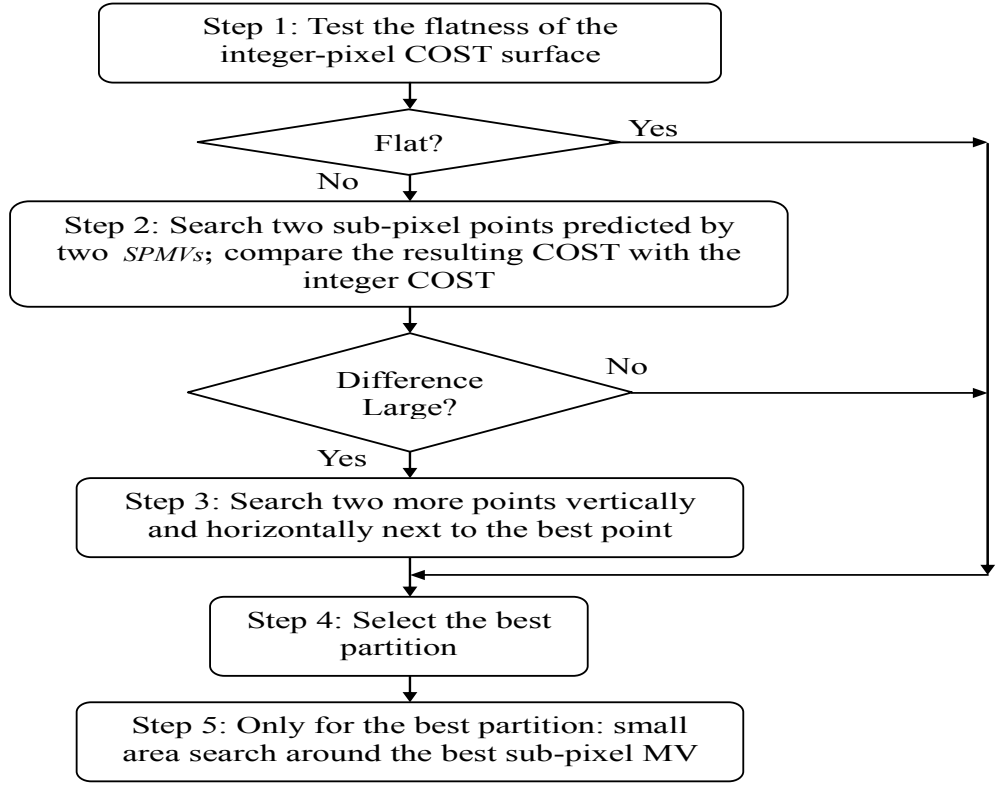

Fig. 2 The proposed Rough-strategy-based Fast Sub-pixel ME algorithm.

## IV. The Fast Sub-pixel ME Algorithm

The entire process of the proposed Rough-strategy-based Fast Sub-pixel Motion Estimation (RFSME) algorithm can be described in Fig. 2. In our algorithm, instead of using only the SAD to model the surface, we use COST [3, 10] as the ME matching cost in the rest of the paper. The COST [3, 10] is defined as:

$$COST = SAD + \lambda_{MOTION} \cdot R(MV) \qquad (5)$$

where $R(MV)$ is the number of bits to code the $MV$, and $\lambda_{MOTION}$ is the Lagrange multiplier [11]. $\lambda_{MOTION}$ is introduced to balance the importance between SAD and $R(MV)$. Note that COST can be viewed as a prediction of the total number of bits for coding both the matching error (i.e., SAD) and the corresponding side information (i.e., $MV$).

In Step 1, the difference between the best COST of the integer position and the two averaged COSTs of its 4 neighboring integer positions (the averaged COST of two vertical neighboring integer positions and the averaged COST of two horizontal neighboring integer positions) are checked. If the difference is small, it means that the COST surface is quite flat, and the best integer COST is close to the optimal sub-pixel COST (and therefore is good enough to estimate the best sub-pixel COST). In this case, the sub-pixel motion estimation is skipped for the current partition. The best COST of the integer position is used in the partition selection in Step 4. The rule for deciding the COST surface flatness is shown in Eqn (6),

$$COST\_Surface = \begin{cases} Not\_Flat & \text{if any of (a),(b),(c) is true} \\ Flat & \text{otherwise} \end{cases} \quad (6)$$

where the conditions (a), (b), and (c) are

(a) $avg\_COST_{vertical} > r_F \cdot COST_{full}$ or $avg\_COST_{horizontal} > r_F \cdot COST_{full}$

(b) if $blocktype(i)$
$\min(|COST_{full} - avg\_COST_{vertical}|, |COST_{full} - avg\_COST_{horizontal}|) > th_1$,

(c) if $blocktype(ii)$
$\min(|COST_{full} - avg\_COST_{vertical}|, |COST_{full} - avg\_COST_{horizontal}|) > th_2$,

$COST_{full}$ is the best COST after full-pixel ME, $avg\_COST_{vertical}$ is the COST average of its two vertical full-pixel neighbors and $avg\_COST_{horizontal}$ is the COST average of its two horizontal full-pixel neighbors. $r_F$ is a ratio parameter to decide whether $avg\_COST_{vertical}$ or $avg\_COST_{horizontal}$ is close to $COST_{full}$. $blocktype(i)$ represents 8×8, 8×4, 4×8, and 4×4 partitions, and $blocktype(ii)$ represents 16×16, 16×8 and 8×16 partitions. $th_1$ and $th_2$ are two thresholds. In the experiment of this paper, $th_1$, $th_2$ and $r_F$ are set to *10*, *20* and *5/4*, respectively. These values are selected based on the experimental statistics.

If the COST surface is not flat in Step 1, in Step 2, two sub-pixel MV prediction methods are used to get two *SPMV*s. The first *SPMV* is calculated by the CBFPS method discussed in Section II, i.e., Eqn (1). The second *SPMV* is calculated by the second-order surface model discussed in Section II. After these two points are searched, the point that has the smallest COST is selected, namely $COST_{step2}$. The motion vector that corresponds to $COST_{step2}$ is defined as $MV_{step2}$.

Table 1 The distribution of absolute distance between the best sub-pixel $MV$ $(x_1, y_1)$ and $MV_{step2}$ $(x_2, y_2)$ (Note: $d=|x_1-x_2|+|y_1-y_2|$ in quarter-pixel units)

| Sequence | d<=0 (%) | d<=1 (%) | d<=2 (%) |
|---|---|---|---|
| News_QCIF | 88.14 | 98.46 | 99.73 |
| Foreman_QCIF | 70.26 | 89.09 | 94.9 |
| Mobile_QCIF | 76.63 | 95.37 | 99.36 |

Table 1 lists the distribution of absolute distance ($d=|x_1-x_2|+|y_1-y_2|$) between the best sub-pixel ($x_1$, $y_1$) $MV$ and ($x_2$, $y_2$) $MV_{step2}$ (the predicted MV corresponding to $COST_{step2}$). The test condition is the same as that described in Section V. It shows that $MV_{step2}$ can provide a good prediction of the best sub-pixel MV. For example, we can see from Table 1 that more than 70% $MV_{step2}$ is exactly the same as the best sub-pixel $MV$ and more than 94% $MV_{step2}$ is within 2 quarter-pixel distance from the best sub-pixel $MV$. Therefore, after Step 2, the assumption is made that $MV_{step2}$ is close to the best sub-pixel MV (but $COST_{step2}$ is not necessarily close to the best sub-pixel COST). The absolute difference between $COST_{step2}$ and the best integer-pixel COST in Step 1 ($COST_{best\_full\_pixel}$) is checked, i.e., $D = |COST_{step2} - COST_{best\_full\_pixel}|$.

If $D$ is small, this means that the COST doesn't decrease much between $COST_{step2}$ and the best integer-pixel COST, and that $COST_{step2}$ is already close to the best sub-pixel COST and is good enough for the mode selection. In this case, $COST_{step2}$ is used in the partition selection in Step 4. The rule for deciding whether $D$ is small or not can be described in Eqn (7),

$$D \quad is \quad \begin{cases} Large & if\ any\ of\ (a),(b),(c)\ is\ true \\ Small & otherwise \end{cases} \quad (7)$$

where

(a) $avg\_COST_{vertical} > r_D \cdot COST_{min\_step2}$ or $avg\_COST_{horizontal} > r_D \cdot COST_{min\_step2}$

(b) if $blocktype(i)$, $D > \frac{1}{2} th_1$

(c) if $blocktype(ii)$, $D > \frac{1}{2} th_2$

$COST_{min\_step2} = min(COST_{step2}, COST_{best\_full\_pixel}$, $avg\_COST_{vertical}$ and $avg\_COST_{horizontal}$ are the same as in Eqn (5). $r_D$ is a ratio parameter to decide whether $avg\_COST_{vertical}$ or $avg\_COST_{horizontal}$ is close to $COST_{min\_step2}$ and it is set to $3/2$ in this paper.

If $D$ is large, $COST_{step2}$ may not be close to the best sub-pixel COST (as shown in Fig. 3(a)). In this case, the two points vertically and the two points horizontally next to $MV_{step2}$ in quarter-pixel resolution will be checked. As shown in Fig. 3(b), the black point is $MV_{step2}$, the grey points are quarter-pixel neighbors of $MV_{step2}$, and the white points are integer neighboring points of $MV_{step2}$. In Step 3, two search points are selected as one point out of $V_1$ and $V_2$, and one point out of $H_1$ and $H_2$. A bilinear model as described below is used to select one of the neighboring points. As shown in Fig. 4 (a), the slopes are first computed (based on Eqn (9)) between the two horizontal neighboring integer points (or the two vertical neighboring integer points) and the best sub-pixel point from Step 2 (the point by $MV_{step2}$). Then, the quarter-pixel neighboring point is selected corresponding to the slope with the *smaller* slope value, as shown in Fig. 4 (b) and Eqn (8).

$$P^{Horizontal}_{step3} = \begin{cases} H_1 & \text{if } S_{H1} < S_{H2} \\ H_2 & \text{if } S_{H1} > S_{H2} \end{cases} \text{ and } P^{Vertical}_{Step3} = \begin{cases} V_1 & \text{if } S_{V1} < S_{V2} \\ V_2 & \text{if } S_{V1} > S_{V2} \end{cases} \quad (8)$$

where

$$S_i = \left| \frac{COST_{integer\_i} - COST_{min\_step2}}{Coord_{integer\_i} - Coord_{min\_step2}} \right|, i = V1, V2, H1, H2 \quad (9)$$

*integer_i* represents the *closest* integer-pixel point in *i*'s direction (i.e. $V_1$ and $V_2$ for the vertical direction and $H_1$ and $H_2$ for the horizontal direction), and *min_step2* represents the best sub-pixel point after Step 2. *Coord* is the coordinate (in quarter-pixel resolution) of the points. The X-coordinate (horizontal direction) is used for $H_1$ and $H_2$, and the Y-coordinate (vertical direction) is used for $V_1$ and $V_2$.

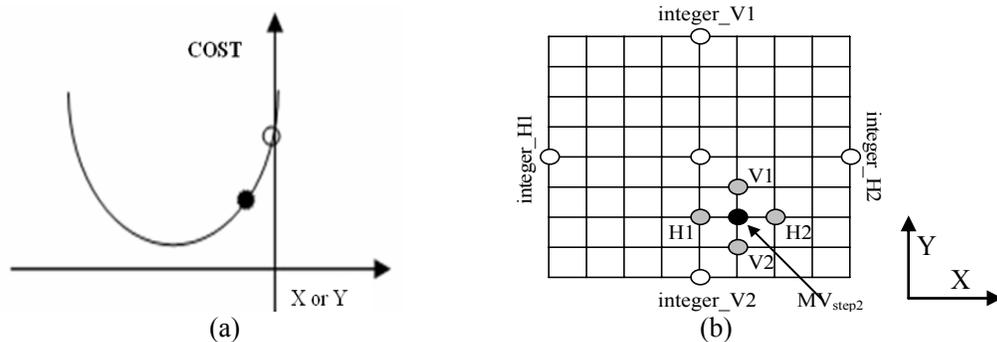

Fig. 3 (a) An example COST surface for $COST_{step2}$ not close to the best sub-pixel COST and, (b) $MV_{step2}$ and its quarter-pixel neighboring points.

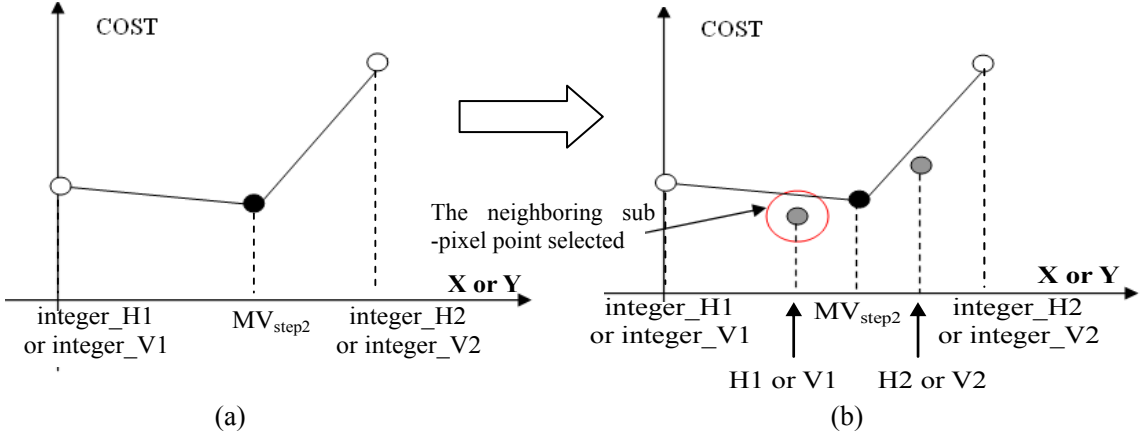

(a)                          (b)

Fig. 4 Using the bilinear model to select neighboring search points (white points: integer pixel; black points: $MV_{step2}$; grey points: neighboring point selected).
(Note: in (a), the left slope is smaller than the right slope. Therefore in (b), the neighboring sub-pixel point on the left is selected).

After Steps 1, 2, and 3, a COST value ($COST_{rough}$) can be obtained for each partition, which is close or equal to the best COST. The sub-pixel MV that corresponds to $COST_{rough}$ is denoted by $MV_{rough}$.

In Step 4, $COST_{rough}$ is used to select the best partition. In Step 5, a small area sub-pixel refinement is performed around $MV_{rough}$. In the proposed algorithm, the eight quarter-pixel neighbors around $MV_{rough}$ are searched. Since Step 5 is performed only for the best partition selected, the average search points per partition is reduced compared to conventional fast sub-pixel search algorithms.

It should be noted that the proposed RFSME algorithm is just one implementation of our idea described in Section III. Our method is general and it could also be implemented in other ways. For example, we can simply skip the sub-pixel search in the 'rough' search step and directly use the best full-pixel searching results to select the partition, and then perform the 'precise' search for the best partition. This can be viewed as a simplified version or extension of the RFSME algorithm.

## V. Experimental Results

We implemented our proposed algorithm on the H.264/AVC reference software JM [5]. In the experiments, each test sequence of 100 frames is coded. The picture coding type is IPPP…, and the frame rate is 30 frames/sec. The search range is 16 for QCIF and 32 for CIF and Standard Definition (SD). The number of reference frames is one. Full search is used for the integer pixel ME in our experiment [5]. It should be noted that our algorithm is general and various other integer pixel ME algorithms can also be easily implemented, as will be discussed later. Six methods are compared for each sequence:

(1) JM Reference Method [5] (Sub-pixel Full Search)

(2) The method in [6] (CBFPS)

(3) The method in [7] (FPME)

(4) The method in [8] (PDFPS)

(5) Use the best Integer COST directly to select the partition and then use JM's method to perform the sub-pixel ME for the best partition (IE+SME-Proposed). As mentioned, this method can be viewed as the simplified version or extension of the proposed RFSME algorithm.

(6) The Proposed RFSME Method (RFSME-Proposed)

In Table 2, the Peak Signal to Noise Ratio (PSNR), bitrate (BR), and average search points (SP) per partition size (SP/PT) [3, 7] for each method are compared for sequences in different resolutions and with different quantization parameters (QPs). The rate-distortion (R-D) curves for some sequences in Table 2 are shown in Fig. 5 (a) and (b). Furthermore, Fig. 5 (c) and (d) shows the BR-SP/PT curves for different methods.

Table 2 Comparison of different ME methods

| Sequence | Method | PSNR (dB) | BR (kbps) | SP/PT |
|---|---|---|---|---|
| Akiyo<br>QCIF (176x144)<br>QP=24 | Full Search | 40.82 | 56.05 | 16 |
| | CBFPS | 40.8 | 57.01 | 6.02 |
| | FPME | 40.81 | 57.12 | 2.92 |
| | PDFPS | 40.79 | 57.26 | 3.30 |
| | IE+SME-Proposed | 40.77 | 60.97 | 0.41 |
| | **RFSME-Proposed** | **40.8** | **57.05** | **0.87** |
| Mobile<br>QCIF (176x144)<br>QP=28 | Full Search | 32.95 | 453.39 | 16 |
| | CBFPS | 32.95 | 453.90 | 7.02 |
| | FPME | 32.95 | 455.82 | 5.81 |
| | PDFPS | 32.95 | 457.17 | 5.72 |
| | IE+SME-Proposed | 32.92 | 484.43 | 0.51 |
| | **RFSME-Proposed** | **32.95** | **456.61** | **3.1** |
| Football CIF (352x288) QP=28 | Full Search | 36.03 | 1440.84 | 16 |
| | CBFPS | 36.01 | 1448.87 | 7.63 |
| | FPME | 36.00 | 1455.60 | 6.21 |
| | PDFPS | 36.01 | 1452.18 | 6.85 |
| | IE+SME-Proposed | 36.01 | 1473.46 | 1.13 |
| | **RFSME-Proposed** | **36.01** | **1451.55** | **3.13** |
| Football CIF (352x288) QP=18 | Full Search | 43.15 | 4456.43 | 16 |
| | CBFPS | 43.15 | 4459.07 | 7.96 |
| | FPME | 43.14 | 4472.68 | 6.46 |
| | PDFPS | 43.14 | 4469.79 | 7.08 |
| | IE+SME-Proposed | 43.14 | 4516.51 | 1.35 |
| | **RFSME-Proposed** | **43.14** | **4462.82** | **3.69** |
| Mobile SD (720x576) QP=28 | Full Search | 33.8 | 8228.28 | 16 |
| | CBFPS | 33.79 | 8253.38 | 7.12 |
| | FPME | 33.78 | 8289.28 | 6.22 |
| | PDFPS | 33.79 | 8302.22 | 6.10 |
| | IE+SME-Proposed | 33.76 | 8625.27 | 1.24 |
| | **RFSME-Proposed** | **33.79** | **8293.79** | **2.88** |
| Flower SD (720x576) QP=24 | Full Search | 37.95 | 8428.84 | 16 |
| | CBFPS | 37.95 | 8432.12 | 6.3 |
| | FPME | 37.94 | 8449.21 | 5.97 |
| | PDFPS | 37.95 | 8461.3 | 5.9 |
| | IE+SME-Proposed | 37.92 | 8631.19 | 0.93 |
| | **RFSME-Proposed** | **37.95** | **8431.03** | **2.39** |

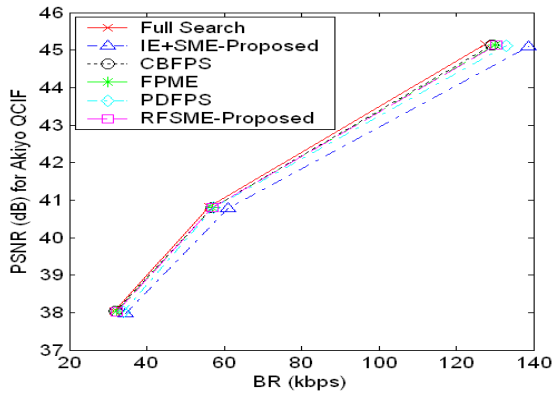
(a) Akiyo QCIF

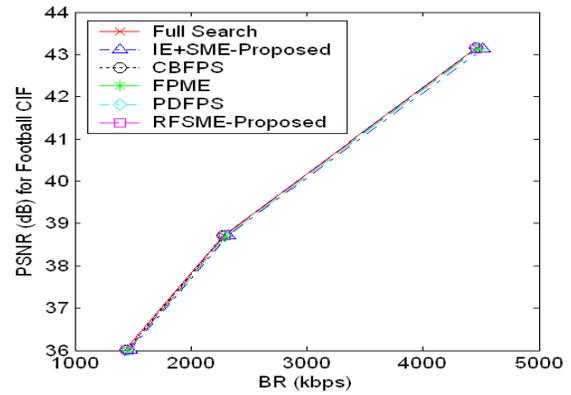
(b) Football CIF

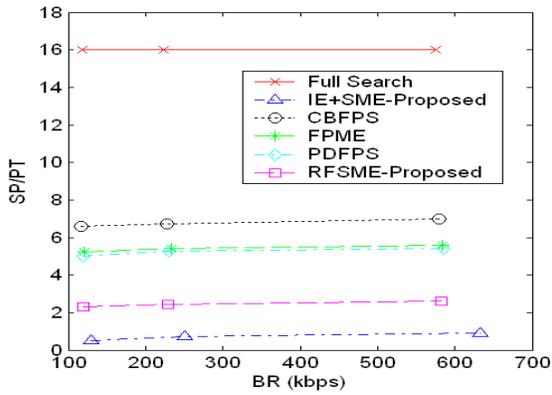
(c) Foreman_QCIF

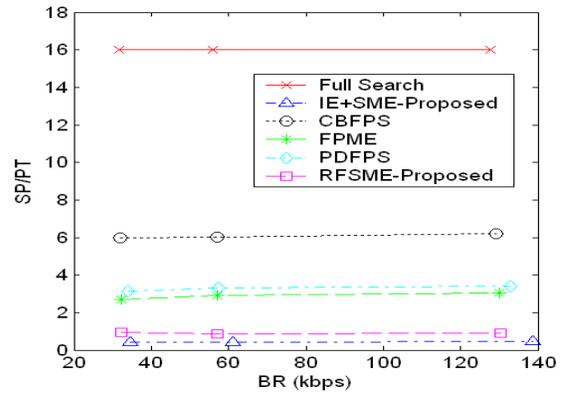
(d) Akiyo_QCIF

Fig. 5 R-D and BR-SP/PT curves comparisons for different methods
((a) and (b): R-D curves, (c) and (d): BR-SP/PT curves)

Several observations can be drawn from Table 2 and Fig. 5:

The previous methods (CBFPS, FPME, and PDFPS) can reduce the SP by reducing the search area around the predicted PMV. However, our proposed methods (IE+SME-proposed and RFSME-proposed) can further reduce more than half the SP compared to previous methods (CBFPS, FPME, and PDFPS) by only performing the 'precise' search on the best partition.

The IE+SME-proposed method can reduce the most number of search points, but the performance decrease is also large for some sequences (e.g. for Akiyo_QCIF in Fig. 5-(a)). This implies that only using the best integer-pixel COST may not always be able to find the best partition mode suitably, and some sub-pixel motion search may be needed to help select the best mode. However, due to its smallest number of SPs, the IE+SME-proposed method can still be very useful in situations where computation complexity is a crucial factor and some quality degradation is tolerable.

The RFSME-proposed method has the best overall performance. Compared with FS and other previous methods (CBFPS, FPME, and PDFPS), the RFSME-proposed method has much smaller SP while keeping almost the same coding performance. Compared with the IE+SME-proposed method, the proposed method has obviously better coding performance. With the RFSME-proposed method, the SP per partition size can be reduced to less than 3 for most sequences. The SP can be further reduced and becomes close to that of the IE+SME-proposed method when coding low motion videos (such as Akiyo_QCIF).

When QP decreases, the SP-per-partition-size for most of the methods will slightly increase. This is because (a) the recovered reference frames are more precise for smaller QPs (i.e., higher PSNR). Therefore, the chance for the MB to select a smaller partition size becomes higher. (b) When the reference frames are more precise, the COST surface for the interpolated sub-pixel locations may become more 'complex', and it may take more steps to find the best sub-pixel location.

Besides the above observations, there are also other advantages of the proposed method. Firstly, the proposed RFSME algorithm models the sub-pixel COST surface based on the 4-neighboring integer COST values. Thus, the algorithm can be easily combined with most of the existing fast integer ME algorithms. Since most fast integer ME algorithms [4, 10, 14-16] (e.g., Simplified Hex

Search [14] and Diamond Search [15]) end the ME process by searching the 4-neighboring points around the best integer point, using the 4-neighbor COST information does not introduce any extra cost to the integer ME process. Furthermore, with the development of new video coding standards (such as High Efficiency Video Coding (HEVC) and Next Generation Video Coding (NGVC) [12]), some existing sub-pixel ME methods may no longer work. For example, with the introduction of Adaptive Interpolation Filter [13] in HEVC or NGVC, the second-order sub-pixel COST surface model may become unsuitable since the interpolation filter will adapt to the frame contents. This will greatly limit the usefulness of many fast sub-pixel methods [7, 9] which rely on this second-order model. Compared to these methods, our proposed methods can still work efficiently after some simple extensions. This is because (a) the basic idea of our method is to reduce sub-pixel SP by performing 'rough' search in the non-best partitions. As long as we can find some way to perform 'rough' search, the proposed method can be easily applied to the new standards. (b) There may be more partition sizes introduced in the future standards. With the introduction of more partition sizes, our proposed method can work even more efficiently by reducing the sub-pixel SP in the non-best partitions.

Table 3 shows another experiment for a multiple reference frame case. In this case, our algorithm first performs the 'rough' search for all partitions on all reference frames and then performs the 'precise' search only for the best partition on the best reference frame. From Table 3, we can see that our algorithm can further reduce SP/PT by performing the 'rough' search on those non-best reference frames.

Table 3 Results for Football CIF using 3 reference frames with QP=28

|          | Full Search | CBFPS   | FPME    | PDFPS   | IE+SME  | RFSME   |
|----------|-------------|---------|---------|---------|---------|---------|
| PSNR (dB)| 36.06       | 36.03   | 36.03   | 36.04   | 36.02   | 36.03   |
| BR (kbps)| 1436.21     | 1442.47 | 1450.04 | 1449.15 | 1478.40 | 1446.45 |
| SP/PT    | 16          | 7.41    | 6.12    | 6.76    | 0.39    | 2.4     |

## VI. Conclusion

In this paper, a fast and effective sub-pixel ME algorithm is proposed. It reduces the number of average SP per partition by more than half compared to conventional fast sub-pixel ME algorithms, with relatively small performance decreases.

## Acknowledgements

The main part of this work was performed when employed at Motorola. This paper is also supported in part by the following grants: Chinese national 973 grants (2010CB731401 and 2010CB731406), National Science Foundation of China grants (60632040, 60902073, 60928003, 60702044 and 60973067).